\documentclass[pre,twocolumn,aps,showpacs]{revtex4}
\newcommand{\bec}[1]{\mbox{\boldmath $ #1$}}
\usepackage{graphicx}
\begin{document}
\title{Excitation of large-scale inertial waves in a rotating
inhomogeneous turbulence}
\author{Tov Elperin}
\email{elperin@menix.bgu.ac.il}
\homepage{http://www.bgu.ac.il/~elperin}
\author{Ilia Golubev}
\email{golubev@bgumail.bgu.ac.il}
\author{Nathan Kleeorin}
\email{nat@menix.bgu.ac.il}
\author{Igor Rogachevskii}
\email{gary@menix.bgu.ac.il}
\homepage{http://www.bgu.ac.il/~gary}
\affiliation{The Pearlstone Center for Aeronautical Engineering
Studies, Department of Mechanical Engineering, Ben-Gurion
University of the Negev, Beer-Sheva 84105, P. O. Box 653, Israel}
\date{\today}
\begin{abstract}
A mechanism of excitation of the large-scale inertial waves in a
rotating inhomogeneous turbulence due to an excitation of a
large-scale instability is found. This instability is caused by a
combined effect of the inhomogeneity of the turbulence and the
uniform mean rotation. The source of the large-scale instability
is the energy of the small-scale turbulence. We determined the
range of parameters at which the large-scale instability occurs,
the growth rate of the instability and the frequency of the
generated large-scale inertial waves.
\end{abstract}


\maketitle

\section{Introduction}

The study of rotating flows is of interest for a wide range of
problems, ranging from engineering (e.g., turbomachinery),
astrophysics (galactic and accretion discs) to geophysics (oceans,
the atmosphere of the Earth, gaseous planets) and weather
predictions (see, e.g., \cite{P87,G90,SF97}). Inertial waves arise
in rotating flows and are observed in the atmosphere of the Earth
and in laboratory rotating flows. In turbulent rotating flows
inertial waves are damped due to a high turbulent viscosity. Thus,
excitation of coherent and undamped inertial waves by turbulence
seems not to be effective. However, large-scale inertial waves are
observed in turbulent rotating flows. A mechanism of excitation of
the large-scale coherent inertial waves in turbulence is not well
understood.

Inertial waves are related with generation of large-scale
vorticity. Generation of a large-scale vorticity in a helical
turbulence due to hydrodynamical $\alpha$ effect was suggested in
\cite{MST83,KMT91,CMP94}. This effect is associated with the
$\alpha \tilde{\bf W}$ term in the equation for the mean
vorticity, where $\tilde{\bf W}$ are the perturbations of the mean
vorticity and $\alpha$ is determined by the hydrodynamical
helicity of turbulent flow. A nonzero hydrodynamical helicity is
caused, e.g., by a combined effect of an uniform rotation and
inhomogeneity of turbulence.

Formation of large-scale vortices in a turbulent rotating flows
was studied experimentally and in numerical simulations (see,
e.g., \cite{ME73,HB82,ZBF97,TYK98,MK99,GL99,AK99}). Formation of
large-scale coherent structures (e.g., large-scale cyclonic and
anticyclonic  vortices) in a small-scale turbulence is one of the
characteristic features of rotating turbulence (see, e.g.,
\cite{ZBF97}). A number of mechanisms have been proposed to
describe generation of a mean flow by a small-scale rotating
turbulence, e.g., the effect of angular momentum mixing
\cite{BT68} and vorticity expulsion \cite{GL68}. The first
experimental demonstration wherein it was shown that the
divergence of the Reynolds stresses can generate an organized mean
circulation was described in \cite{ZBF97}.

There is a certain similarity between mean rotation and a mean
velocity shear. Generation of a mean vorticity in a nonhelical
homogeneous incompressible turbulent flow with an imposed mean
velocity shear due to an excitation of a large-scale instability
was studied in \cite{EKR03}. This instability is caused by a
combined effect of the large-scale shear motions (''skew-induced"
deflection of equilibrium mean vorticity) and ''Reynolds
stress-induced" generation of perturbations of the mean vorticity.
This instability and the dynamics of the mean vorticity are
associated with the Prandtl's turbulent secondary flows (see,
e.g., \cite{P52,T56,B87,C94}). However, a turbulence with an
imposed mean velocity shear and a uniformly rotating turbulence
are different. In particular, the mean vorticity is generated by a
homogeneous nonhelical sheared turbulence \cite{EKR03}. On the
other hand, the mean vorticity cannot be generated by a
homogeneous uniformly rotating nonhelical turbulence (see below).
The main difference between these two flows is that the mean
velocity shear produces work in a turbulent flow, while a uniform
rotation does not produce work in a homogeneous turbulent flow.

There are other interesting problems related with the inertial
waves including, e.g., the effect of inertial waves on the onset
of convection and on the turbulence dynamics. In particular, the
onset of convection in the form of inertial waves in a rotating
fluid sphere were studied in \cite{BS04}. On the other hand, the
modification of turbulence dynamics by rotation is due to the
presence of small-scale inertial waves in rotating flows (see,
e.g., \cite{G90,CS99,BPS03,GA03,CR04}).

The main goal of this paper is to study large-scale structures
formed in a rotating inhomogeneous turbulence. In particular, we
investigate the excitation of large-scale inertial waves. These
structures are associated with a generation of a large-scale
vorticity due to the excitation of the large-scale instability in
an uniformly rotating inhomogeneous turbulence. The excitation of
the mean vorticity in this system requires an inhomogeneity of
turbulence.

This paper is organized as follows. In Section II we formulated
the governing equations, the assumptions and the procedure of the
derivation. In Section III the effective force was determined,
which allowed us to derive the mean-filed equations and to study
the excitation of large-scale inertial waves in Section IV. The
large-scale instability was investigated in Section IV
analytically for a weakly inhomogeneous turbulence and numerically
for an arbitrary inhomogeneous turbulence. Conclusions and
applications of the obtained results are discussed in Section V.
In Appendixes A, B and C the detailed derivation of the effective
force is performed.

\section{The governing equations}

The system of equations for the evolution of the velocity ${\bf
v}$ and vorticity ${\bf W} \equiv \bec{\nabla} {\bf \times} {\bf
v}$ reads:
\begin{eqnarray}
\biggl[{\partial \over \partial t} + {\bf v} \cdot \bec{\nabla}
\biggr]{\bf v} = - {\bec{\nabla} P \over \rho} + 2 \, {\bf v} {\bf
\times} {\bf \Omega} + \nu \Delta {\bf v} + {\bf F}_{\rm st} \;,
\label{L1} \\
{\partial {\bf W} \over \partial t} = \bec{\nabla} {\bf \times} \,
({\bf v} {\bf \times} {\bf W} + 2 \, {\bf v} {\bf \times} {\bf
\Omega} - \nu \bec{\nabla} {\bf \times} {\bf W}) \;, \label{B1}
\end{eqnarray}
where ${\bf v}$ is the fluid velocity with $\bec{\nabla} \cdot
{\bf v} = 0 ,$ $ \, P $ is the fluid pressure, $\, {\bf F}_{\rm
st}$ is an external stirring force with a zero mean value, ${\bf
\Omega}$ is a constant angular velocity and $ \nu$ is the
kinematic viscosity. Equation~(\ref{B1}) follows from the
Navier-Stokes equation~(\ref{L1}). We use a mean field approach
whereby the velocity, pressure and vorticity are separated into
the mean and fluctuating parts: $ {\bf v} = \bar{\bf U} + {\bf u}
,$ $ P = \bar{P} + p $ and $ {\bf W} = \bar{\bf W} + {\bf w} ,$
the fluctuating parts have zero mean values, $ \bar{\bf U} =
\langle {\bf v} \rangle ,$ $ \, \bar{P} = \langle P \rangle $ and
$ \bar{{\bf W}} = \langle {\bf W} \rangle .$ Averaging
Eqs.~(\ref{L1}) and~(\ref{B1}) over an ensemble of fluctuations we
obtain the equations for the mean velocity $\bar{\bf U}$ and mean
vorticity $\bar{{\bf W}} :$
\begin{eqnarray}
\biggl[{\partial \over \partial t} + \bar{\bf U} \cdot
\bec{\nabla} \biggr] \bar{\bf U} = - {\bec{\nabla} \bar{P} \over
\rho} + 2 \, \bar{\bf U} {\bf \times} {\bf \Omega} + \bec{\cal F}
+ \nu \Delta \bar{\bf U}  \;,
\label{L2} \\
{\partial \bar{{\bf W}} \over \partial t} = \bec{\nabla} {\bf
\times} \, (\bar{\bf U} {\bf \times} \bar{{\bf W}} + 2 \, \bar{\bf
U} {\bf \times} {\bf \Omega} + \langle {\bf u} {\bf \times} {\bf
w} \rangle - \nu \bec{\nabla} {\bf \times} \bar{{\bf W}})  \;,
\label{B2}
\end{eqnarray}
where ${\cal F}_i = - \nabla_j \langle u_i u_j \rangle .$ Note
that the effect of turbulence on the mean vorticity is determined
by the Reynolds stresses $\langle u_i u_j \rangle$ because
\begin{eqnarray}
\langle {\bf u} {\bf \times} {\bf w} \rangle_i = - \nabla_j
\langle u_i u_j \rangle + {1 \over 2} \nabla_i \langle {\bf u}^2
\rangle \; .
\label{B3}
\end{eqnarray}

Consider a steady state solution of Eqs.~(\ref{L2}) and~(\ref{B2})
in the form: $ \bar{\bf U}^{(s)} = 0 $ and   $\bar{\bf W}^{(s)} =
0 .$ In order to study a stability of this equilibrium we consider
perturbations of the mean velocity, $\tilde{\bf U} ,$ and the mean
vorticity, $\tilde{{\bf W}} .$ The linearized equations for the
small perturbations of the mean velocity and the mean vorticity
are given by
\begin{eqnarray}
{\partial \tilde{\bf U} \over \partial t} = - {\bec{\nabla}
\tilde{P} \over \rho} + 2 \, \tilde{\bf U} {\bf \times} {\bf
\Omega} + \tilde{\bec{\cal F}}(\tilde{\bf U}) + \nu \Delta
\tilde{\bf U}  \;,
\label{L3} \\
{\partial \tilde{\bf W} \over \partial t} = \bec{\nabla} {\bf
\times} \, (2 \, \tilde{\bf U} {\bf \times} \bec{\Omega} +
\tilde{\bec{\cal F}}(\tilde{\bf U}) - \nu \bec{\nabla} {\bf
\times} \tilde{{\bf W}}) \;, \label{B4}
\end{eqnarray}
where $ \tilde{\bec{\cal F}}(\tilde{\bf U}) = - \nabla_j (f_{ij} -
f_{ij}^{(0)}) $ is the effective force, $ f_{ij} = \langle u_i u_j
\rangle $ and $f_{ij}^{(0)}$ is the second moment of the velocity
field in a background turbulence (with a zero gradient of the mean
velocity). Thus, the mean fields $\tilde{\bf U}$ and $\tilde{\bf
W}$ represent deviations from the equilibrium solution $ \bar{\bf
U}^{(s)} = 0 $ and $\bar{\bf W}^{(s)} = 0 .$ This equilibrium
solution is a steady state solution of Eqs.~(\ref{L2})
and~(\ref{B2}). Note that the characteristic times and spatial
scales of small-scale fluctuations of velocity and vorticity ${\bf
u}$ and ${\bf w}$ are much smaller than that of the mean fields
$\tilde{\bf U}$ and $\tilde{\bf W}$.

In order to obtain a closed system of equations in the next
Section we derived an equation for the effective force $
\tilde{\bec{\cal F}} .$

\section{The effective force}

In this section we derive an equation for the effective force $
\tilde{\bec{\cal F}} .$ The mean velocity gradients can affect
turbulence. The reason is that additional essentially
non-isotropic velocity fluctuations can be generated by tangling
of the mean-velocity gradients with the Kolmogorov-type
turbulence. The source of energy of this "tangling turbulence" is
the energy of the Kolmogorov turbulence. The tangling turbulence
was introduced by Wheelon \cite{W57} and Batchelor et al.
\cite{BH59} for a passive scalar and by Golitsyn \cite{G60} and
Moffatt \cite{M61} for a passive vector (magnetic field).
Anisotropic fluctuations of a passive scalar (e.g., the number
density of particles or temperature) are generated by tangling of
gradients of the mean passive scalar field with a random velocity
field. Similarly, anisotropic magnetic fluctuations are excited by
tangling of the mean magnetic field with the velocity
fluctuations. The Reynolds stresses in a turbulent flow with mean
velocity gradients is another example of a tangling turbulence.
Indeed, they are strongly anisotropic in the presence of mean
velocity gradients and have a steeper spectrum $ (\propto
k^{-7/3}) $ than a Kolmogorov turbulence (see, e.g.,
\cite{L67,WC72,SV94,IY02,EKRZ02}). The anisotropic velocity
fluctuations of tangling turbulence were studied first by Lumley
\cite{L67}.

To derive an equation for the effective force $ \tilde{\bec{\cal
F}} $ we use equation for fluctuations $ {\bf u}(t,{\bf r}) $
which is obtained by subtracting Eq.~(\ref{L2}) for the mean field
from Eq.~(\ref{L1}) for the total field:
\begin{eqnarray}
{\partial {\bf u} \over \partial t} &=& - (\bar{\bf U} \cdot
\bec{\nabla}) {\bf u} - ({\bf u} \cdot \bec{\nabla}) \bar{\bf U} -
{\bec{\nabla} p \over \rho}
\nonumber\\
& & + 2 \, {\bf u} {\bf \times} \bec{\Omega} + {\bf F}_{\rm st} +
{\bf U}^{N} \;, \label{B5}
\end{eqnarray}
where
\begin{eqnarray}
{\bf U}^{N} &=& \langle ({\bf u} \cdot \bec{\nabla}) {\bf u}
\rangle - ({\bf u} \cdot \bec{\nabla}) {\bf u} + \nu \Delta {\bf
u} \; .
\label{R10}
\end{eqnarray}
We consider a turbulent flow with large Reynolds numbers $ ({\rm
Re} = l_{0} u_{0} / \nu \gg 1) $, where $ u_{0} $ is the
characteristic velocity in the maximum scale $ l_{0} $ of
turbulent motions. We assume that there is a separation of scales,
i.e., the maximum scale of turbulent motions $ l_0 $ is much
smaller then the characteristic scale of inhomogeneities of the
mean fields. Using Eq.~(\ref{B5}) we derived equation for the
second moment of turbulent velocity field $ f_{ij}({\bf k, R})
\equiv \int \langle u_i ({\bf k} + {\bf K} / 2) u_j(-{\bf k} +
{\bf K} / 2) \rangle \exp(i {\bf K} {\bf \cdot} {\bf R}) \,d {\bf
K} $:
\begin{eqnarray}
{\partial f_{ij}({\bf k, R}) \over \partial t} &=& {\cal G}_{ijmn}
f_{mn} + F_{ij} + f_{ij}^{(N)} \; \label{A6}
\end{eqnarray}
(see Appendix A), where ${\cal G}_{ijmn} = I_{ijmn}(\tilde{\bf U})
+ N_{ijmn}({\bf \Omega}) ,$
\begin{eqnarray}
I_{ijmn}(\tilde{\bf U}) &=& \biggl[2 k_{iq} \delta_{mp}
\delta_{jn} + 2 k_{jq} \delta_{im} \delta_{pn} - \delta_{im}
\delta_{jq} \delta_{np}
\nonumber\\
&& - \delta_{iq} \delta_{jn} \delta_{mp} + \delta_{im} \delta_{jn}
k_{q} {\partial \over \partial k_{p}} \biggr] \nabla_{p} \tilde
U_{q} \;,
\label{A14}\\
N_{ijmn}({\bf \Omega}) &=& 2 \, \Omega_q k_{pq} (\varepsilon_{imp}
\, \delta_{nj} + \varepsilon_{jmp} \, \delta_{ni}) \;, \label{L4}
\end{eqnarray}
and $ {\bf R} $ and $ {\bf K} $ correspond to the large scales,
and $ {\bf r} $ and $ {\bf k} $ to the small scales (see Appendix
A), $ \delta_{ij} $ is the Kronecker tensor, $ k_{ij} = k_i  k_j /
k^2 ,$ and  $ \bec{\nabla} = \partial /
\partial {\bf R} ,$ $\, F_{ij}({\bf k},{\bf R}) = \langle \tilde F_i
({\bf k},{\bf R}) u_j(-{\bf k},{\bf R}) \rangle + \langle u_i({\bf
k},{\bf R}) \tilde F_j(-{\bf k },{\bf R}) \rangle $, $\, {\bf
\tilde F} ({\bf k},{\bf R},t) = - {\bf k} {\bf \times} ({\bf k}
{\bf \times} {\bf F}_{\rm st} ({\bf k},{\bf R})) / k^2 $ and $
f_{ij}^{(N)}({\bf k},{\bf R}) $ are the terms which are related
with the third moments appearing due to the nonlinear terms.  The
third moments terms $f_{ij}^{(N)}$ are defined as
\begin{eqnarray*}
f_{ij}^{(N)}({\bf k},{\bf R}) &=& \langle P_{in}({\bf k}_1) \hat
U^{N}_{n}({\bf k}_1) u_j({\bf k}_2) \rangle
\nonumber\\
&& + \langle u_i({\bf k}_1) P_{jn}({\bf k}_2) \hat U^{N}_{n}({\bf
k}_2) \rangle \;,
\end{eqnarray*}
where $\hat {\bf U}^{N}_{n}({\bf k})$ is the Fourier transform of
${\bf U}^{N}$ determined by Eq.~(\ref{R10}), ${\bf k}_1 = {\bf k}
+ {\bf K} / 2$, $\, {\bf k}_2 = -{\bf k} + {\bf K} / 2 $ and $
P_{ij}({\bf k}) = \delta_{ij} - k_{ij} $.

Equation~(\ref{A6}) is written in a frame moving with a local
velocity $ \tilde{\bf U} $ of the mean flow. In Eq.~(\ref{A6}) for
the second moments of the turbulent velocity field we neglected
small terms $\sim O(\nabla^2)$ , where the terms with $\nabla \sim
O(L^{-1})$ contain the large-scale spatial derivatives. These
terms are of the order of $(l_0/L)^2$, where the maximum scale of
turbulent motions $l_0$ is much smaller than the vertical size of
the turbulent region $L$.

Equation~(\ref{A6}) for the background turbulence (with a zero
gradient of the mean fluid velocity $ \nabla_{i} \tilde U_{j} = 0
$ and ${\bf \Omega}=0)$ reads
\begin{eqnarray}
{\partial f_{ij}^{(0)}({\bf k, R}) \over \partial t} = F_{ij} +
f_{ij}^{(N,0)} \;, \label{B7}
\end{eqnarray}
where the superscript $ {(0)} $ corresponds to the background
turbulence, and we assumed  that the tensor $F_{ij}({\bf k, R}) ,$
which is determined by a stirring force, is independent of the
mean velocity gradients and of a constant mean angular velocity.
Equation for the deviations $f_{ij} - f_{ij}^{(0)}$ from the
background turbulence is given by
\begin{eqnarray}
{\partial (f_{ij} - f_{ij}^{(0)}) \over \partial t} = {\cal
G}_{ijmn} f_{mn} + f_{ij}^{(N)} - f_{ij}^{(N,0)} \; . \label{B8}
\end{eqnarray}

Equation~(\ref{B8}) for the deviations of the second moments
$f_{ij} - f_{ij}^{(0)} $ in ${\bf k}$-space contains the
deviations of the third moments $f_{ij}^{(N)} - f_{ij}^{(N,0)}$
and a problem of closing the equations for the higher moments
arises. Various approximate methods have been proposed for the
solution of problems of this type (see, {\em e.g.,}
\cite{MY75,O70,Mc90}). The simplest procedure is the $ \tau $
approximation which was widely used for study of different
problems of turbulent transport (see, e.g.,
\cite{O70,PFL76,KRR90,RK2000,BK04}). One of the simplest
procedures, that allows us to express the deviations of the terms
with the third moments $f_{ij}^{(N)} - f_{ij}^{(N,0)}$ in ${\bf
k}$-space in terms of that for the second moments $f_{ij} -
f_{ij}^{(0)} ,$ reads
\begin{eqnarray}
f_{ij}^{(N)} - f_{ij}^{(N,0)} &=& - {f_{ij} - f_{ij}^{(0)} \over
\tau (k)} \;, \label{A1}
\end{eqnarray}
where $ \tau (k) $ is the scale dependent correlation time of the
turbulent velocity field. Here we assumed that the time $ \tau(k)
$ is independent of the mean velocity gradients (for a weak mean
velocity shear). We considered also the case of slow rotation
rate. In this case a modification of the correlation time of fully
developed turbulence by slow rotation is small. This allows us to
suggest that Eq.~(\ref{A1}) is valid for a slow rotation rate.

The $ \tau $-approximation  is different from Eddy Damped Quasi
Normal Markowian (EDQNM) approximation. A principle difference
between these two approaches is as follows (see \cite{O70,Mc90}).
The EDQNM closures do not relax to the equilibrium, and this
procedure does not describe properly the motions in the
equilibrium state. In EDQNM approximation, there is no dynamically
determined relaxation time, and no slightly perturbed steady state
can be approached \cite{O70}. In the $ \tau $-approximation, the
relaxation time for small departures from equilibrium is
determined by the random motions in the equilibrium state, but not
by the departure from equilibrium \cite{O70}. Analysis performed
in \cite{O70} showed that the $ \tau $-approximation describes the
relaxation to the equilibrium state (the background turbulence)
more accurately than the EDQNM approach.

Note that we applied the $ \tau $-approximation (\ref{A1}) only to
study the deviations from the background turbulence which are
caused by the spatial derivatives of the mean velocity and a
uniform rotation. The background turbulence is assumed to be
known. Here we use the following model for the background
isotropic and weakly inhomogeneous turbulence:
\begin{eqnarray}
f^{(0)}_{ij}({\bf k},{\bf R}) &=& \frac{{\cal E}(k)}{8 \pi k^{2}}
\biggl[P_{ij}({\bf k}) + \frac{i}{2 k^2}(k_i \nabla_j
\nonumber\\
&&- k_j \nabla_i) \biggr] \langle {\bf u}^2 \rangle^{(0)} \;
\label{A2}
\end{eqnarray}
(see e.g., \cite{EKR95}), where $ P_{ij}({\bf k}) = \delta_{ij} -
k_{ij} $, $\, \delta_{ij} $ is the Kronecker tensor and $ k_{ij} =
k_i  k_j / k^2 $, $\, \tau(k) = 2 \tau_0 \bar \tau(k) ,$ $ \,
{\cal E}(k) = - d \bar \tau(k) / dk ,$ $ \, \bar \tau(k) = (k /
k_{0})^{1-q} ,$ $ \, 1 < q < 3 $  is the exponent of the kinetic
energy spectrum (e.g., $ q = 5/3 $ for Kolmogorov spectrum), $
k_{0} = 1 / l_{0} .$

The mean velocity gradient $ \nabla_i \tilde{\bf U} $ causes
generation of anisotropic velocity fluctuations (tangling
turbulence). Equations~(\ref{B8})-(\ref{A2}) allow to determine
the second moment $ f_{ij}({\bf R}) = \int f_{ij}({\bf k},{\bf R})
\, d{\bf k} $:
\begin{eqnarray}
f_{ij}({\bf R}) &=& f_{ij}^{(0)}({\bf R}) - \nu_{_{T}}^{(0)}
M_{ij} - {1 \over 6} l_0^2 \, \Omega S_{ij} \; \label{B15}
\end{eqnarray}
(see Appendix B), where $\nu_{_{T}}^{(0)} = \tau_0 \langle {\bf
u}^2 \rangle^{(0)} / 6 $ and the tensors $M_{ij} $ and $S_{ij}$
are determined in Appendix B. The definition of the function
$\nu_{_{T}}^{(0)}$ yields $\langle {\bf u}^2 \rangle^{(0)}({\bf
R}) = 6 \nu_{_{T}}^{(0)}({\bf R}) / \tau_0$. Since we assumed that
$\tau_0$ is independent of ${\bf R}$, the spatial profile of the
function $\nu_{_{T}}^{(0)}$ [e.g., given by Eq.~(\ref{PB1}) in
Sect. IV-B] determines the spatial profile of $\langle {\bf u}^2
\rangle^{(0)}$. Equation~(\ref{B15}) allows to determine the
effective force $ \tilde{\cal F}_i = - {\nabla}_j [ f_{ij}({\bf
R}) - f_{ij}^{(0)}({\bf R})]:$
\begin{eqnarray}
\tilde{\cal F}_i = \nu_{_{T}}^{(0)} (M_{ij} \Lambda_j + \nabla_j
M_{ij}) + {1 \over 6} l_0^2 \, \Omega (S_{ij} \Lambda_j + \nabla_j
S_{ij}) \;,
\nonumber\\
\label{B14}
\end{eqnarray}
where $ {\bf \Lambda} = (\bec{\nabla} l_0^2) /  l_0^2 =
(\bec{\nabla} \nu_{_{T}}^{(0)}) / \nu_{_{T}}^{(0)} .$

Note that when $ \Lambda = const ,$ the effective force $
\tilde{\bec{\cal F}} $ does not have the term $ \propto \alpha
\tilde{\bf W} ,$ where $\alpha$ describes the hydrodynamic
$\alpha$-effect.  The hydrodynamic $\alpha$-effect was introduced
in the equation for the mean vorticity (see, e.g.,
\cite{MST83,KMT91,CMP94}), similarly to the $\alpha$-effect in the
equation for the evolution of the mean magnetic field (see, e.g.,
\cite{M78}). The reason for the absence of the $ \alpha \tilde{\bf
W} $ term in $ \tilde{\bec{\cal F}} $ is as follows. Let us
suggest the opposite, i.e., that $ \tilde{\bec{\cal F}} \propto
\alpha \tilde{\bf W} = \alpha \bec{\nabla} {\bf \times} \tilde{\bf
U} .$ Since the effective force $ \tilde{\cal F}_i = - {\nabla}_j
[ f_{ij}({\bf R}) - f_{ij}^{(0)}({\bf R})], $ we obtain
\begin{eqnarray}
f_{ij}({\bf R}) - f_{ij}^{(0)}({\bf R}) \propto - \alpha
\varepsilon_{ijk} \tilde{U}_k \; . \label{R1}
\end{eqnarray}
Here we used the identity $ \tilde{W}_i = \varepsilon_{ijk}
\nabla_j \tilde{U}_k$ and we took into account that when $ \Lambda
= const ,$ the hydrodynamic $\alpha$ is constant. Note also that
in our paper we considered incompressible velocity field. The
condition~(\ref{R1}) is in contradiction with the Galilean
invariance, because the Reynolds stresses in the considered case
may depend on the gradient of the mean velocity field rather than
on the mean velocity itself. When $ \Lambda$ is not constant, the
effective force $ \tilde{\bec{\cal F}} $ can have the term $
\propto \alpha \tilde{\bf W} .$ However, this effect is not in the
scope of our paper (e.g., this case cannot be described in the
framework of the gradient approximation).

\section{The large-scale instability in an inhomogeneous
turbulence}

\begin{figure}
\centering
\includegraphics[width=8cm]{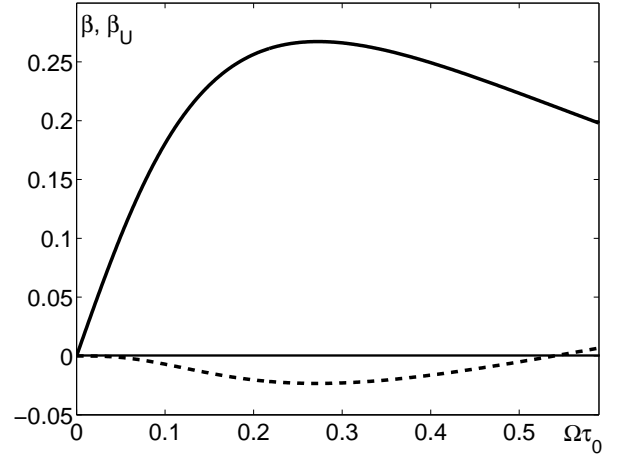}
\caption{\label{FIG1} The rotation rate dependence of the
functions $ \beta(\Omega \tau_0)$ (solid) and $
\beta_{_{U}}(\Omega \tau_0)$ (dashed).}
\end{figure}

\begin{figure}
\centering
\includegraphics[width=8cm]{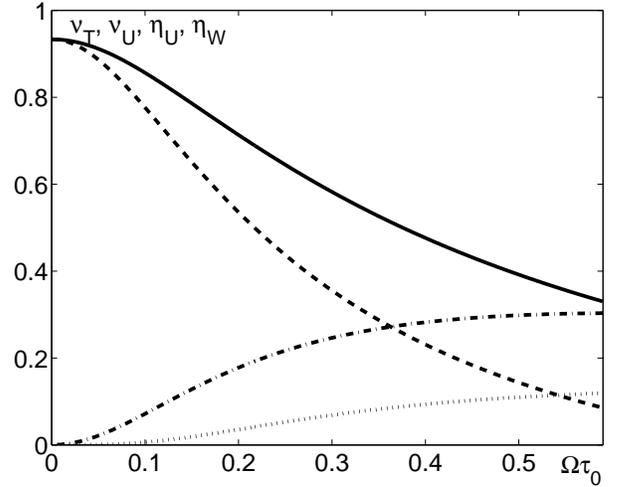}
\caption{\label{FIG2} The rotation rate dependence of the
functions $\nu_{_{T}}(\Omega \tau_0)$ (solid), $ \nu_{_{U}}(\Omega
\tau_0) $ (dashed-dotted), $\eta_{_{W}}(\Omega \tau_0)$ (dashed)
and $\eta_{_{U}}(\Omega \tau_0)$ (dotted).}
\end{figure}

For simplicity we consider the case when the turbulence is
inhomogeneous along the rotation axis, i.e.,  $ {\bf \Lambda} =
\Lambda(z) {\bf e}_z ,$ $ \, {\bf \Omega} = \Omega {\bf e}_z .$
After calculating $ [\bec{\nabla} {\bf \times} (\bec{\nabla} {\bf
\times} \tilde{\bf U})]_z$ from Eq.~(\ref{L3}) and $\tilde{W}_z$
from Eq.~(\ref{B4}) we arrive at the following equations written
in nondimensional form
\begin{eqnarray}
\Delta {\partial \tilde{U}_z \over \partial t}   &=& - [\hat G -
\beta_{_{U}} \, \Delta_\perp \nabla_z ] \, \tilde{W}_z +
[\nu_{_{T}} \, \Delta^2 + \nu_{_{T}} \, \Lambda \, \Delta \,
\nabla_z
\nonumber\\
&& + \, \nu_{_{U}} \, \Lambda \, \Delta_\perp \, \nabla_z +
\eta_{_{U}} \, \Delta_\perp \nabla_z^2 ] \tilde{U}_z \;,
\label{E2}\\
{\partial \tilde{W}_z \over \partial t} &=& (\hat G - \beta \,
\Lambda \, \Delta_\perp) \tilde{U}_z + [\nu_{_{T}} \, \nabla_z^2 +
\nu_{_{T}} \, \Lambda \, \nabla_z
\nonumber\\
&& + \, \eta_{_{W}} \, \Delta_\perp] \tilde{W}_z \;, \label{E3}
\end{eqnarray}
where $\Delta = \Delta_\perp + \nabla_z^2$, $\, \nabla_z =
\partial / \partial z$,
\begin{eqnarray*}
\hat G = 2 \, a_\ast \, \nabla_z + \beta [\Lambda \, \nabla_z^2 +
\Delta \nabla_z] \;,
\end{eqnarray*}
\begin{eqnarray*}
\beta(\Omega, z) &=& \nu_{_{T}}^{(0)}(z) (\omega  / 8)
[D_5(\omega) / 2 - D_6(\omega)] \;,
\\
\beta_{_{U}}(\Omega, z) &=& \nu_{_{T}}^{(0)}(z) (\omega  / 8)
D_6(\omega) \;,
\\
\nu_{_{T}}(\Omega, z) &=& \nu_{_{T}}^{(0)}(z) [D_1(\omega) / 2 +
D_2(\omega)] \;,
\\
\nu_{_{U}}(\Omega, z) &=& \, \eta_{_{U}}(\Omega, z) + 2
\nu_{_{T}}^{(0)}(z) D_2(\omega)  \;,
\\
\eta_{_{U}}(\Omega, z) &=& \nu_{_{T}}^{(0)}(z) D_4(\omega) \;,
\\
\eta_{_{W}}(\Omega, z) &=& \nu_{_{T}}^{(0)}(z) [D_1(\omega) / 2 -
D_3(\omega)]  \; .
\end{eqnarray*}
Here $\omega = 8 \tau_0 \Omega ,$ $\, a_\ast = \Omega L^{2} /
\nu_{_{T}}^\ast \gg 1 $, and we used Eq.~(\ref{B14}). The
functions $D_k(\omega)$ are determined in Appendix B.

In Eqs.~(\ref{E2}) and~(\ref{E3}) we use the following
dimensionless variables: length is measured in units of $L$, time
in units of $ L^{2} / \nu_{_{T}}^\ast ,$ the function $\Lambda(z)$
is measured in the units of $L^{-1}$, the function
$\nu_{_{T}}^{(0)}(z)$ is measured in the units of
$\nu_{_{T}}^\ast$, the perturbations of velocity $\tilde{U}_z$ and
vorticity $\tilde{W}_z$ are measured in units of $ U_\ast $ and $
U_\ast / L  ,$ respectively. The functions $\, \beta(\Omega
\tau_0)$, $\, \beta_{_{U}}(\Omega \tau_0)$, $\, \nu_{_{T}}(\Omega
\tau_0)$, $\, \nu_{_{U}}(\Omega \tau_0) $, $\, \eta_{_{W}}(\Omega
\tau_0)$ and $\, \eta_{_{U}}(\Omega \tau_0)$ are shown in FIGS.
1-2. All these functions shown in FIGS. 1-2 are normalized by
$\nu_{_{T}}^{(0)}(z)$, e.g, $\nu_{_{T}}(\Omega \tau_0) \equiv
\nu_{_{T}}(\Omega, z) / \nu_{_{T}}^{(0)}(z)$, and similarly for
other functions.

\subsection{The weakly inhomogeneous turbulence}

Assume that functions $\nu_{_{T}}^{(0)}(z)$ and $\Lambda(z)$ vary
slowly with $z$ in comparison with the variations of the mean
velocity $\tilde{U}_z(z) $ and mean vorticity $\tilde{W}_z(z) $.
Let us seek for the solution of Eqs.~(\ref{E2}) and~(\ref{E3}) in
the form $ \propto \exp(\gamma t - i {\bf K} \cdot {\bf R}) .$ Let
us first consider perturbations with the wave numbers $ K^2_\perp
\ll K^2_z .$ Since $ \bec{\nabla} \cdot \tilde{\bf U} = 0 ,$ the
velocity components $\tilde{U}_z \ll |\tilde{\bf U}_\perp| .$ Thus
the growth rate of the inertial waves with the frequency
\begin{eqnarray}
\omega_w =  - {\rm sgn}(\beta \,\Lambda) \, {2 \, a_\ast \, K_z
\over K} \;, \label{WE6}
\end{eqnarray}
is given by
\begin{eqnarray}
\gamma_w &=& |\beta(\Omega \tau_0) \,\Lambda \, K_z|  -
\nu_{_{T}}(\Omega \tau_0) K^2 \;, \label{E6}
\end{eqnarray}
where $\gamma = \gamma_w + i \omega_w ,$ the wave number $K$ is
measured in the units of $L^{-1}$ and $\gamma$ is measured in the
units of $\nu_{_{T}}^\ast / L^{2} .$ The maximum growth rate of
the inertial waves, $ \gamma_m = [\beta(\Omega \tau_0) \,
\Lambda]^2 / 4 \nu_{_{T}}(\Omega \tau_0) ,$ is attained at $ K =
K_m = |\beta(\Omega \tau_0) \, \Lambda| / 2 \nu_{_{T}}(\Omega
\tau_0) .$ For a very small rotation rate, i.e., for $ \omega
\equiv 8 \Omega \tau_0 \ll 1 ,$ the turbulent viscosity $
\nu_{_{T}}(\Omega \tau_0) \approx (q + 3) / 5 $ and $\beta(\Omega
\tau_0) \approx (32/15) \Omega \tau_0 ,$ where the parameter $q$
is the exponent of the kinetic energy spectrum of the background
isotropic and weakly inhomogeneous turbulence (e.g., $ q = 5/3 $
for Kolmogorov spectrum), and this parameter varies in the range
$1 < q < 3$. Note that the inertial waves are helical, i.e., the
large-scale hydrodynamic helicity of the motions in the inertial
waves is $ \tilde{\bf U} \cdot (\bec{\nabla} {\bf \times}
\tilde{\bf U}) = 2 |\tilde{\bf U}_\perp|^2 K_z \not= 0 .$ This
instability is caused by a combined effect of the inhomogeneity of
the turbulence and the uniform mean rotation [see the first term
in Eq.~(\ref{E6})].

Now we consider the opposite case, i.e., the perturbations with
the wave numbers $ K^2_\perp \gg K^2_z .$ Since $ \bec{\nabla}
\cdot \tilde{\bf U} = 0 ,$ the velocity components $\tilde{U}_z
\gg |\tilde{\bf U}_\perp| .$ When $K^2 \, |\beta(\Omega) \,
\Lambda| / 4 a_\ast \ll K_z \ll K_\perp $, the growth rate of
perturbations with the frequency
\begin{eqnarray}
\omega_w =  {\rm sgn}(\beta \, \Lambda) \, {2 \, a_\ast \, K_z
\over K} \;, \label{TWE6}
\end{eqnarray}
is given by
\begin{eqnarray}
\gamma_w &=& {1 \over 2} \biggl[ |\beta(\Omega \tau_0) \, \Lambda|
\, K - [\nu_{_{T}}(\Omega \tau_0) + \eta_{_{W}}(\Omega \tau_0)] \,
K^2 \biggr] .
\nonumber\\
\label{TE6}
\end{eqnarray}
The maximum growth rate of perturbations, $\gamma_m =
[\beta(\Omega \tau_0) \, \Lambda]^2 / 8 [\nu_{_{T}}(\Omega \tau_0)
+ \eta_{_{W}}(\Omega \tau_0)] ,$ is attained at $ K = K_m =
|\beta(\Omega \tau_0) \, \Lambda| / 2[\nu_{_{T}}(\Omega \tau_0) +
\eta_{_{W}}(\Omega \tau_0)] .$ This case corresponds to  $ a_\ast
\gg \beta^2(\Omega \tau_0) / 4 \nu_{_{T}}(\Omega \tau_0) .$ The
large-scale hydrodynamic helicity of the flow is $ \tilde{\bf U}
\cdot (\bec{\nabla} {\bf \times} \tilde{\bf U}) = 4 K_\perp
|\tilde{\bf U}_z|^2 \, {\rm sgn}(\beta) \not= 0 .$

\subsection{Numerical results}

In this Section we take into account the inhomogeneity of the
functions $\nu_{_{T}}^{(0)}(z)$ and $\Lambda(z)$. We introduce a
new variable $V = \Delta \tilde{U}_z$ and consider an eigenvalue
problem for a system of Eqs.~(\ref{E2}) and~(\ref{E3}). We seek
for a solution of Eqs.~(\ref{E2}) and~(\ref{E3}) in the form $
\propto \Psi(z) \exp(\gamma t) J_0(K_\perp \, r) ,$ where $J_0(x)$
is the Bessel function of the first kind. After the substitution
of this solution into Eqs.~(\ref{E2}) and~(\ref{E3}) we obtain the
system of the ordinary differential equations which is solved
numerically.

We used the cylindrical geometry $(z,r,\phi)$ with $z$-axis along
the rotation axis and consider the axisymmetric solution (i.e.,
there are no derivatives with respect to the polar angle $\phi$).
The turbulence is inhomogeneous along the rotation axis. We use
the periodic boundary conditions in $z$ direction for
Eqs.~(\ref{E2}) and~(\ref{E3}), i.e., $\tilde{U}_z(z=0,r) =
\tilde{U}_z(z=L,r)$, $\tilde{U}_z'(z=0,r) = \tilde{U}_z'(z=L,r)$,
$\tilde{U}_z''(z=0,r) = \tilde{U}_z''(z=L,r)$,
$\tilde{U}_z'''(z=0,r) = \tilde{U}_z'''(z=L,r)$,
$\tilde{W}_z(z=0,r) = \tilde{W}_z(z=L,r)$ and $\tilde{W}_z'(z=0,r)
= \tilde{W}_z'(z=L,r) $, where $\tilde{U}' = \nabla_z \tilde{U}$.
We also use the condition $\tilde{U}_r(z,r=0) =\tilde{U}_r(z,r=R)
= 0$, where $R$ is the radius of the turbulent region.

We have chosen the vertical profile of the function
$\nu_{_{T}}^{(0)}(z)$ in the following form
\begin{eqnarray}
\nu_{_{T}}^{(0)}(z) &=& 1 - C \, \biggl\{ 1 - \exp \biggl[- 2
\Lambda_0^2 \biggl(\frac{z}{L}- \frac{1}{2} \biggr)^2 \biggr]
\biggr\}\;,
\nonumber\\
\label{PB1}\\
C &=& {1 - \nu_{_{T}}^{(b)} \over 1 - \exp (- \Lambda_0^2 / 2)}
\;,
\nonumber
\end{eqnarray}
with two values of the parameter $\nu_{_{T}}^{(b)} =$ $\, 0.1$ and
$0.3$; and two values of the parameter $ \Lambda_0=$ $\,12$ and
$30$. The vertical profile of the turbulent viscosity
$\nu_{_{T}}^{(0)}(z)$ is shown in FIG.~3. The maximum of
turbulence intensity is located at $z=L/2$. The form of the chosen
spatial profile of the function $\nu_{_{T}}^{(0)}(z)$ is simple
enough and universal. It allows us to vary the size of the region
occupied by turbulence (by changing the parameter $\Lambda_0$) and
the difference in the level of the turbulence between the center
and boundary of the region (by changing the parameter
$\nu_{_{T}}^{(b)}$), i.e., it allows us to change the
inhomogeneity of the turbulence. The numerical solution of Eqs.
~(\ref{E2}) and~(\ref{E3}) was performed also for other spatial
profiles of the function $\nu_{_{T}}^{(0)}(z)$. However, the final
results do not depend strongly on the details in the spatial
profile of the function $\nu_{_{T}}^{(0)}(z)$. Note also that the
chosen spatial profile of the function $\nu_{_{T}}^{(0)}(z)$ can
mimic the distribution of turbulence in galactic and accretion
discs (see, e.g., \cite{RSS88}).

\begin{figure}
\centering
\includegraphics[width=8cm]{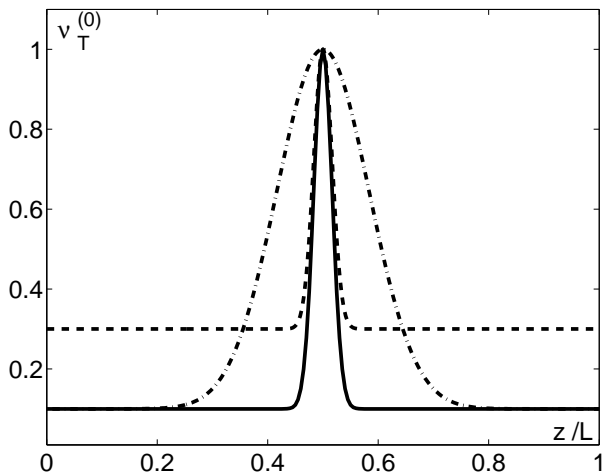}
\caption{\label{FIG3} The vertical profile of the turbulent
viscosity $\nu_{_{T}}^{(0)}(z)$ for $\Lambda_0 = 30$ and
$\nu_{_{T}}^{(b)}=0.1$ (solid); $\Lambda_0 = 30$ and $\,
\nu_{_{T}}^{(b)}=0.3$ (dashed); $\Lambda_0 = 12$ and $\,
\nu_{_{T}}^{(b)}=0.1$ (dashed-dotted).}
\end{figure}

\begin{figure}
\centering
\includegraphics[width=8cm]{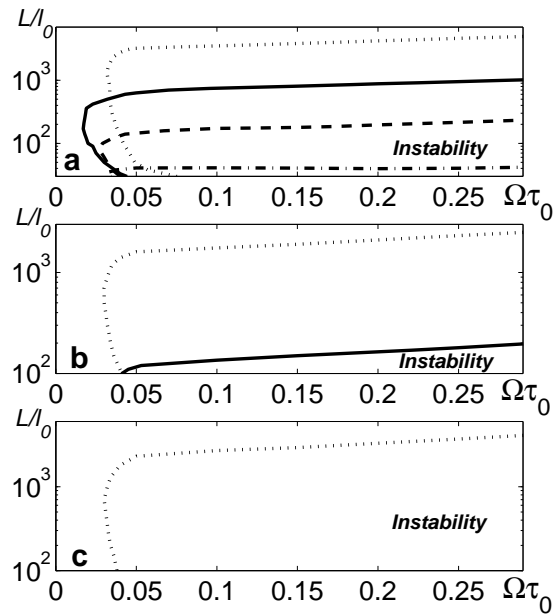}
\caption{\label{FIG4} The range of parameters $(L /l_0 , \Omega
\tau_0)$ for which the large-scale instability occurs $ (\gamma_w
> 0) $ for {\bf (a)}. $\Lambda_0 = 12$, $\, \nu_{_{T}}^{(b)}=0.1$;
{\bf (b)}. $\Lambda_0 = 30$, $\, \nu_{_{T}}^{(b)}=0.1$; {\bf (c)}.
$\Lambda_0 = 30$, $\, \nu_{_{T}}^{(b)}=0.3$; and different values
of the parameter $\mu$: $\, \, \mu=0.1$ (dotted), $\mu=0.5$
(solid), $\mu=1$ (dashed), $\mu=2$ (dashed-dotted).}
\end{figure}

\begin{figure}
\centering
\includegraphics[width=8cm]{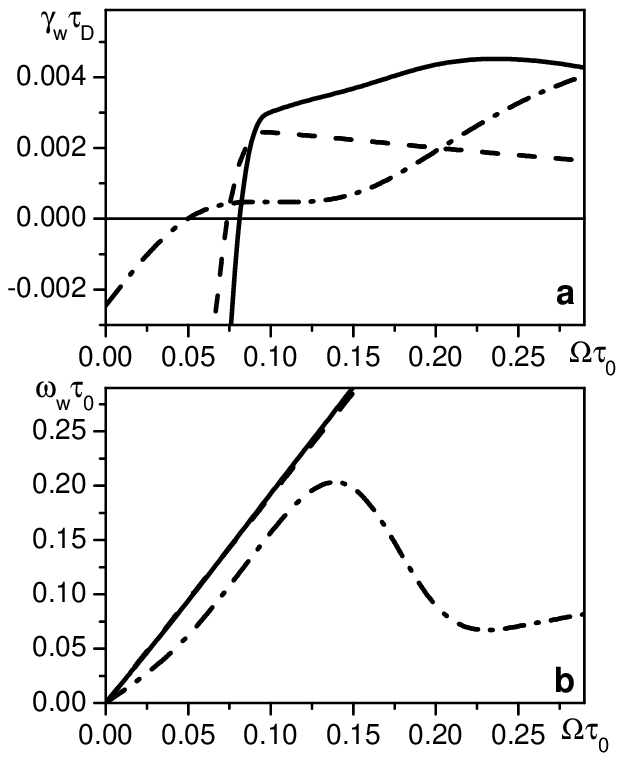}
\caption{\label{FIG5} The rotation rate $(\Omega \tau_0)$
dependencies of {\bf (a)} the growth rate $\gamma_w \tau_{_{D}}$
of the large-scale instability and {\bf (b)} the frequency
$\omega_w \tau_0$ of the generated waves due to the large-scale
instability for $\Lambda_0 = 12$, $\, \nu_{_{T}}^{(b)}=0.1$, $\,
\mu=0.1$ and different values of the parameter $L /l_0$: $\, \,
L/l_0 = 50$ (solid), $\, \, L/l_0 = 100$ (dashed) and $\, \, L/l_0
= 500$ (dashed-dotted). Here $ \tau_{_{D}} = L^{2} /
\nu_{_{T}}^\ast .$}
\end{figure}

\begin{figure}
\centering
\includegraphics[width=8cm]{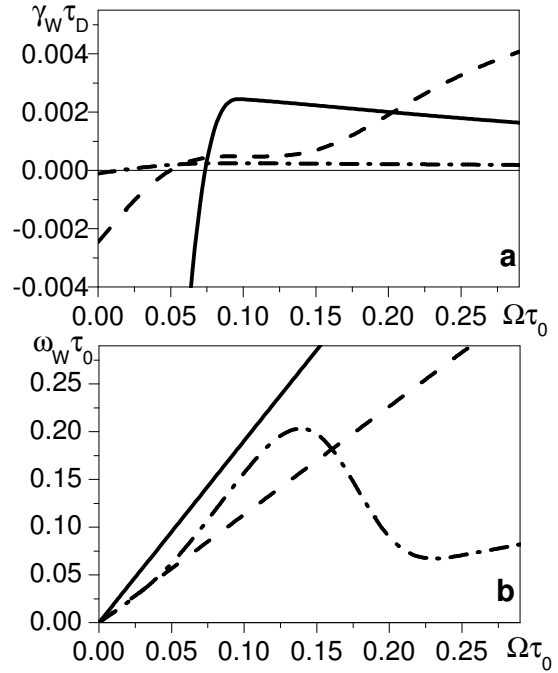}
\caption{\label{FIG6} The rotation rate $(\Omega \tau_0)$
dependencies of {\bf (a)} the growth rate $\gamma_w \tau_{_{D}}$
of the large-scale instability and {\bf (b)} the frequency
$\omega_w \tau_0$ of the generated waves due to the large-scale
instability for $\Lambda_0 = 30$, $\, \nu_{_{T}}^{(b)}=0.1$, $\,
\mu=0.1$ and different values of the parameter $L /l_0$: $\, \,
L/l_0 = 100$ (solid), $\, \, L/l_0 = 500$ (dashed) and $\, \,
L/l_0 = 1000$ (dashed-dotted).}
\end{figure}

The sufficient condition for the excitation of the instability is
$ \gamma_w > 0 $. The range of parameters $ (L /l_0 , \Omega
\tau_0) $ for which the large-scale instability occurs is shown in
FIG.~4 for different values of the parameters $\mu$,
$\nu_{_{T}}^{(0)}(z)$ and $\Lambda_0$. Here $\mu=L /L_r $, $\, L$
is the vertical size of the whole region, $L_r$ is a radius from
the center of the structure at which the energy $\tilde{U}_r^2$ of
the radial velocity perturbations is maximum. Note that the
maximum radial (horizontal) size $R$ of the whole region is of the
order of $\sim 4L_r$. The decrease of the parameter $\mu$ causes
increase of the range of the large-scale instability. On the other
hand, the increase of the size of the highly intense turbulent
region (i.e., decrease of the parameter $\Lambda_0)$ results in
the increase of the range of the instability.

The rotation rate dependencies of the growth rate $\gamma_w
\tau_{_{D}}$ of the large-scale instability and the frequency
$\omega_w \tau_0$ of the generated waves due to the large-scale
instability are shown in FIGS. 5-6, where $\tau_{_{D}} = L^{2} /
\nu_{_{T}}^\ast $. There is a threshold in the rotation rate for
the large-scale instability: $\Omega_\ast \tau_0 \approx 0.025$,
and when $\Omega > \Omega_\ast$, the instability is excited. The
instability threshold in the parameter $L$ is $L > 10 l_0$.

Note that the characteristic time $(\sim 2 \pi / \gamma_w)$ of the
growth of perturbations of the mean fields $\tilde{\bf U}$ and
$\tilde{\bf W}$ is by 5 orders of magnitudes larger than the
turbulent correlation time $\tau_0$. The period of oscillations $
T = 2 \pi / \omega_w$ of inertial waves is at least 10 times
larger than the turbulent correlation time $\tau_0$. The minimum
value of the period of rotation $ T_R = 2 \pi / \Omega$ is at
least 20 times larger than the turbulent correlation time
$\tau_0$. All spatial scales: the vertical size of the turbulent
region,  $L ,$ and the vertical size of the highly intense
turbulence $(\sim L / \Lambda_0)$ are much larger than the maximum
scale of turbulent motions $l_0$ (e.g., $L / \Lambda_0$ is at
least 10 times larger than the maximum scale of turbulent motions
$l_0$). This implies that there is indeed a separation of scales
as we assumed in the derivations. Note also that the range of
validity of the obtained results is $\tau_0 \ll T$ for a
statistically stationary background turbulence. In particular, we
assumed that the characteristic time of evolution of the second
moments is much smaller than the turbulent correlation time. The
asymptotic formulas~(\ref{WE6})-(\ref{TE6}) are in agreement with
the obtained numerical results.

It must be noted that turbulent flow with an imposed mean linear
velocity shear and uniformly rotating flows are essentially
different. In particular, in a turbulent flow with an imposed mean
linear velocity shear there are no waves similar to the inertial
waves which exist in a uniformly rotating flows. The reason is
that any shear motions have a nonzero symmetric part $ (\partial
\hat U)_{ij} $ of the gradient of the mean velocity, where $
(\partial \hat U)_{ij} = (\nabla_i \bar U_{j} + \nabla_j \bar
U_{i}) / 2 .$ In addition, the difference between these two flows
is that the mean velocity shear produces work in a turbulent flow,
while a uniform rotation does not produce work in homogeneous
turbulent flow.

\section{Conclusions}

We studied formation of large-scale structures in a rotating
inhomogeneous nonhelical turbulence. We found a mechanism for the
excitation of the large-scale inertial waves which is associated
with a generation of a large-scale vorticity due to the excitation
of the large-scale instability in a uniformly rotating
inhomogeneous turbulence. It was shown that the mean vorticity
cannot be generated by a homogeneous uniformly rotating nonhelical
turbulence. The excitation of the mean vorticity in this flow
requires also an inhomogeneity of turbulence. Therefore, the
large-scale instability is caused by a combined effect of the
inhomogeneity of the turbulence and the uniform mean rotation. The
source of the large-scale instability is the energy of the
small-scale turbulence. The rotation and inhomogeneity of
turbulence provide a mechanism for transport of energy from
turbulence to large-scale motions. We determined the range of
parameters at which the large-scale instability occurs.

Some of the results obtained in this study, e.g., the expression
for the effective force in a homogeneous turbulence, are in
compliance  with the previous studies of rotating turbulence
\cite{G03} [see Appendix B, Eq.~(\ref{AB15})]. It is plausible to
suggest that the results of recent experiments \cite{FA04} can be
explained by the large-scale instability discussed in this paper.

\begin{acknowledgments}
This work was partially supported by The German-Israeli Project
Cooperation (DIP) administered by the Federal Ministry of
Education and Research (BMBF) and by the Israel Science Foundation
governed by the Israeli Academy of Science.
\end{acknowledgments}

\appendix

\section{Derivation of Eq.~(\ref{A6})}

In order to derive Eq.~(\ref{A6}) we use a mean field approach,
i.e., a correlation function is written as follows
\begin{eqnarray*}
\langle u_i({\bf x}) u_j ({\bf  y}) \rangle &=& \int \langle u_i
({\bf k}_1) u_j ({\bf k}_2) \rangle  \exp[i({\bf  k}_1 {\bf \cdot}
{\bf x} \\
&& + {\bf k}_2 {\bf \cdot} {\bf y})] \,d{\bf k}_1 \, d{\bf k}_2
\\
&=& \int f_{ij}({\bf k, R}) \exp(i {\bf k} {\bf \cdot} {\bf r})
\,d {\bf k} \;,
\\
f_{ij}({\bf k, R}) &=& \int \langle u_i ({\bf k} + {\bf  K} / 2)
u_j(-{\bf k} + {\bf  K} / 2) \rangle
\\
&& \times \exp(i {\bf K} {\bf \cdot} {\bf R}) \,d {\bf K} \;
\end{eqnarray*}
(see, {\em e.g.,} \cite{RS75,KR94}), where $ {\bf R} $ and $ {\bf
K} $ correspond to the large scales, and $ {\bf r} $ and $ {\bf k}
$ to the small scales, {\em i.e.,} $ {\bf R} = ({\bf x} +  {\bf
y}) / 2  ,$ $ \quad {\bf r} = {\bf x} - {\bf y},$ $ \quad {\bf K}
= {\bf k}_1 + {\bf k}_2,$ $ \quad {\bf k} = ({\bf k}_1 - {\bf
k}_2) / 2 .$ This implies that we assumed that there exists a
separation of scales, i.e., the maximum scale of turbulent motions
$ l_0 $ is much smaller then the characteristic scale of
inhomogeneities of the mean fields.

Now we calculate
\begin{eqnarray}
{\partial f_{ij}({\bf k}_1,{\bf k}_2) \over \partial t} &\equiv&
\langle P_{in}({\bf k}_1) {\partial u_{n}({\bf k}_1) \over
\partial t} u_j({\bf k}_2) \rangle
\nonumber\\
&& + \langle u_i({\bf k}_1) P_{jn}({\bf k}_2) {\partial u_{n}({\bf
k}_2) \over \partial t} \rangle \;, \label{A8}
\end{eqnarray}
where we multiplied equation of motion (\ref{B5}) rewritten in $
{\bf k} $-space by $ P_{ij}({\bf k}) = \delta_{ij} - k_{ij} $ in
order to exclude the pressure term from the equation of motion, $
\delta_{ij} $ is the Kronecker tensor and $ k_{ij} = k_i  k_j /
k^2 .$ This yields the equation for $ f_{ij}({\bf k, R}) $ [see
Eq.~(\ref{A6})]. For the derivation of Eq.~(\ref{A6}) we used the
following equation
\begin{eqnarray}
&& i k_i \int f_{ij}({\bf k} - {1 \over 2}{\bf  Q}, {\bf K} - {\bf
Q}) \tilde U_{p}({\bf  Q}) \exp(i {\bf K} {\bf \cdot} {\bf R}) \,d
{\bf  K} \,d {\bf  Q}
\nonumber\\
& & = -{1 \over 2} \tilde U_{p} \nabla _i f_{ij} + {1 \over 2}
f_{ij} \nabla _i \tilde U_{p} -  {i \over 4} (\nabla _s \tilde
U_{p}) \biggl(\nabla _i {\partial f_{ij} \over \partial k_s}
\biggr)
\nonumber\\
& & +  {i \over 4} \biggl( {\partial f_{ij} \over
\partial k_s} \biggr) (\nabla _s \nabla _i \tilde U_{p}) \; .
\label{D1}
\end{eqnarray}
To derive Eq.~(\ref{D1}) we multiply  the equation
$\bec{\nabla}~\cdot~{\bf u} = 0 ,$ written in ${\bf k}$-space for
$u_i({\bf k}_1 - {\bf Q}) ,$ by $u_j({\bf k}_2) \tilde U_{p}({\bf
Q}) \exp(i {\bf K} {\bf \cdot} {\bf R}) ,$ and integrate over
${\bf K}$ and ${\bf Q}$, and average over the ensemble of velocity
fluctuations. Here ${\bf k}_1 = {\bf k} + {\bf  K} / 2$ and ${\bf
k}_2 = -{\bf k} + {\bf K} / 2 .$ This yields
\begin{eqnarray}
&& \int i \biggl(k_i + {1 \over 2} K_i - Q_i \biggr) \langle
u_i({\bf k} + {1 \over 2}{\bf K} - {\bf Q}) u_j(-{\bf k} + {1
\over 2}{\bf K}) \rangle
\nonumber\\
& & \times \tilde U_{p}({\bf  Q}) \exp{(i {\bf K} {\bf \cdot} {\bf
R})} \,d {\bf  K} \,d {\bf Q} = 0  \; . \label{D2}
\end{eqnarray}
Next, we introduce new variables: $ \tilde {\bf k}_{1} = {\bf k} +
{\bf  K} / 2 - {\bf  Q} ,$ $ \tilde {\bf k}_{2} = - {\bf k} + {\bf
K} / 2 $ and $ \tilde {\bf k} = (\tilde {\bf k}_{1} - \tilde {\bf
k}_{2}) / 2 = {\bf k} - {\bf  Q} / 2,$ $ \tilde {\bf K} = \tilde
{\bf k}_{1} + \tilde {\bf k}_{2} = {\bf  K} - {\bf  Q} .$ This
allows us to rewrite Eq.~(\ref{D2}) in the form
\begin{eqnarray}
& & \int i \biggl(k_i + {1 \over 2} K_i - Q_i \biggr) f_{ij}({\bf
k} - {1 \over 2}{\bf Q}, {\bf K} - {\bf Q}) \tilde U_{p}({\bf  Q})
\nonumber\\
& & \times \exp{(i {\bf K} {\bf \cdot} {\bf R})} \,d {\bf  K} \,d
{\bf Q} = 0  \; .
\label{D3}
\end{eqnarray}
Since $ |{\bf Q}| \ll |{\bf k}| $ we can use the Taylor series
expansion
\begin{eqnarray}
f_{ij}({\bf k} - {\bf Q}/2, {\bf  K} - {\bf  Q}) \simeq
f_{ij}({\bf k},{\bf  K} - {\bf  Q})
\nonumber\\
- \frac{1}{2} {\partial f_{ij}({\bf k},{\bf  K} - {\bf Q}) \over
\partial k_s} Q_s  + O({\bf Q}^2) \; .
\label{D4}
\end{eqnarray}
We also use the following identities:
\begin{eqnarray}
&& [f_{ij}({\bf k},{\bf R}) \tilde U_{p}({\bf R})]_{\bf  K} = \int
f_{ij}({\bf k},{\bf  K} - {\bf  Q}) \tilde U_{p}({\bf Q}) \,d {\bf
Q} \;,
\nonumber \\
&& \nabla_{p} [f_{ij}({\bf k},{\bf R}) \tilde U_{p}({\bf R})] =
\int i K_{p} [f_{ij}({\bf k},{\bf R}) \tilde U_{p}({\bf R})]_{\bf
K}
\nonumber\\
&& \times \exp{(i {\bf K} {\bf \cdot} {\bf R})} \,d {\bf  K}  \;,
\label{D5}
\end{eqnarray}
where $[f_{ij}({\bf k},{\bf R}) \tilde U_{p}({\bf R})]_{\bf K}$
denotes a Fourier transformation. Therefore, Eqs.
(\ref{D3})-(\ref{D5}) yield Eq.~(\ref{D1}).

\section{The Reynolds stresses}

In this Appendix we derive equation for the Reynolds stresses
using Eq.~(\ref{B8}). We assume that the characteristic time of
variation of the second moment $f_{ij}({\bf k},{\bf R})$ is
substantially larger than the correlation time $\tau(k)$ for all
turbulence scales. Thus in a steady-state Eq.~(\ref{B8}) reads
\begin{eqnarray}
[\hat L({\bf \Omega}) - \tau(k) \, \hat I(\tilde{\bf U})] \, (\hat
f &-& \hat f^{(0)}) = \tau(k) \, [\hat N({\bf \Omega})
\nonumber\\
&& + \hat I(\tilde{\bf U})] \, \hat f^{(0)} \;, \label{B9}
\end{eqnarray}
where we used Eq.~(\ref{A1}). Hereafter we use the following
notations: $\hat f \equiv f_{ij}({\bf k, R}) ,$ $\, \hat f^{(N)}
\equiv f_{ij}^{(N)}({\bf k, R}) ,$ $\, \hat f^{(0)} \equiv
f_{ij}^{(0)}({\bf k, R}) ,$ $ \, \hat I(\tilde{\bf U}) \hat f
\equiv I_{ijmn}(\tilde{\bf U}) f_{mn}({\bf k, R}) $ and $ \, \hat
N({\bf \Omega}) \hat f \equiv N_{ijmn}({\bf \Omega}) f_{mn}({\bf
k, R}) ,$ and $\hat L({\bf \Omega}) \equiv L_{ijmn}({\bf \Omega})
= \delta_{im} \delta_{jn} - \tau(k) \, N_{ijmn}({\bf \Omega})$.
Multiplying Eq.~(\ref{B9}) by the inverse operator $ \hat
L^{-1}({\bf \Omega})$ yields
\begin{eqnarray}
[\hat E &-& \tau(k) \, \hat L^{-1}({\bf \Omega}) \, \hat
I(\tilde{\bf U})] \, (\hat f - \hat f^{(0)}) = - [\hat E - \hat
L^{-1}({\bf \Omega})
\nonumber\\
&& - \tau(k) \, \hat L^{-1}({\bf \Omega}) \, \hat I(\tilde{\bf
U})] \, \hat f^{(0)} \;, \label{L5}
\end{eqnarray}
where $ \hat E \equiv \delta_{im} \delta_{jn} $ and we used an
identity
\begin{eqnarray*}
\hat E - \hat L^{-1}({\bf \Omega}) = - \tau(k) \, \hat L^{-1}({\bf
\Omega}) \, \hat N({\bf \Omega})  \; .
\end{eqnarray*}
The latter identity follows from the definition: $ \hat
L^{-1}({\bf \Omega}) \hat L({\bf \Omega}) = \hat E .$ The inverse
operator $ \hat L^{-1}({\bf \Omega}) $ is given by
\begin{eqnarray}
&& \hat L^{-1}({\bf \Omega}) \equiv L_{ijmn}^{-1}({\bf \Omega}) =
{1 \over 2} [B_1 \, \delta_{im} \delta_{jn} + B_2 \, k_{ijmn}
\nonumber\\
&& + B_3 \, (\varepsilon_{ipm} \delta_{jn} + \varepsilon_{jpn}
\delta_{im}) \hat k_p + B_4 \, (\delta_{im} k_{jn} + \delta_{jn}
k_{im})
\nonumber\\
&& + B_5 \, \varepsilon_{ipm} \varepsilon_{jqn} k_{pq} + B_6 \,
(\varepsilon_{ipm} k_{jpn} + \varepsilon_{jpn} k_{ipm}) ] \;,
\label{L6}
\end{eqnarray}
where $B_1 = 1 + \chi(2 \psi) ,$ $\, B_2 = B_1 + 2 - 4 \chi(\psi)
,$ $\, B_3 = 2 \psi \, \chi(2 \psi) ,$ $\, B_4 = 2 \chi(\psi) -
B_1 ,$ $\, B_5 = 2 - B_1  ,$ $\, B_6 = 2 \psi \, [\chi(\psi) -
\chi(2 \psi)] ,$ $\, \chi(x) = 1 / (1 + x^2) $ and $ \psi = 2
\tau(k) \, ({\bf k}~\cdot~{\bf \Omega}) / k .$

Multiplying Eq.~(\ref{L5}) by the operator $ \hat E + \tau(k) \,
\hat L^{-1}({\bf \Omega}) \, \hat I(\tilde{\bf U})$ yields the
second moment $ \hat f \equiv f_{ij}({\bf k},{\bf R})$:
\begin{eqnarray}
\hat f &\approx& [\hat L^{-1}({\bf \Omega}) + \tau(k) \, \hat
L^{-1}({\bf \Omega}) \, \hat I(\tilde{\bf U}) \, \hat L^{-1}({\bf
\Omega})] \, \hat f^{(0)} ,
\nonumber\\
\label{B11}
\end{eqnarray}
where we neglected terms which are of the order of $O(|\nabla
\tilde{\bf U}|^2) .$ Since $ L^{-1}_{ijmn}({\bf \Omega}) \,
P_{mn}({\bf k}) = P_{ij}({\bf k})$, Eq.~(\ref{B11}) reads
\begin{eqnarray}
\hat f &\approx& \hat f^{(0)} + \tau(k) \, \hat L^{-1}({\bf
\Omega}) \, \hat I(\tilde{\bf U}) \, \hat f^{(0)} \; . \label{B12}
\end{eqnarray}
The first term in Eq.~(\ref{B12}) describes the background
turbulence. The second term in Eq.~(\ref{B12}) determines effects
of both, rotation and mean gradients of the velocity perturbations
on the turbulence. The integration in ${\bf k}$-space yields the
second moment $ \tilde f_{ij}({\bf R}) = \int \tilde f_{ij}({\bf
k},{\bf R}) \, d{\bf k} $ which is determined by Eq.~(\ref{B15}),
where we used the notation $\tilde f \equiv \tilde f_{ij} = f_{ij}
- f_{ij}^{(0)} ,$ and the tensors $M_{ij} $ and $S_{ij}$ are given
by
\begin{eqnarray}
M_{ij} &=& D_1(\omega) \, ({\partial \tilde U})_{ij} + D_2(\omega)
\, Q_{ij} + D_3(\omega) \, T_{ij}
\nonumber\\
&& + D_4(\omega) \, (\bec{\hat{\omega}} \cdot \bec{\nabla})
(\bec{\hat{\omega}} \cdot \tilde{\bf U}) \, \omega_{ij} \;,
\label{B16} \\
S_{ij} &=& D_5(\omega) \, K_{ij} + D_6(\omega) \, R_{ij} \;,
\label{B19}
\end{eqnarray}
where
\begin{eqnarray}
Q_{ij} &=& (\hat \omega_{i} \nabla_j + \hat \omega_j \nabla_i)
(\bec{\hat{\omega}} \cdot \tilde{\bf U})
\nonumber\\
&& + (\bec{\hat{\omega}} \cdot \bec{\nabla}) (\hat \omega_{i}
\tilde U_j + \hat \omega_j \tilde U_i) \;,
\label{B17}\\
T_{ij} &=& (\bec{\hat{\omega}} {\bf \times} \bec{\nabla})_i
(\bec{\hat{\omega}} {\bf \times} \tilde{\bf U})_j +
(\bec{\hat{\omega}} {\bf \times} \bec{\nabla})_j
(\bec{\hat{\omega}} {\bf \times} \tilde{\bf U})_i \;,
\label{B18}\\
K_{ij} &=& \hat \omega_n [\varepsilon_{imn} ({\partial \tilde
U})_{mj} + \varepsilon_{jmn} ({\partial \tilde U})_{mi}] \;,
\label{B20}\\
R_{ij} &=& [\hat \omega_i (\bec{\hat{\omega}} {\bf \times}
\bec{\nabla})_j + \hat \omega_j (\bec{\hat{\omega}} {\bf \times}
\bec{\nabla})_i] (\bec{\hat{\omega}} \cdot \tilde{\bf U})
\nonumber\\
&& + (\bec{\hat{\omega}} \cdot \bec{\nabla}) [\hat \omega_i
(\bec{\hat{\omega}} {\bf \times} \tilde{\bf U})_j + \hat \omega_j
(\bec{\hat{\omega}} {\bf \times} \tilde{\bf U})_i] \;,
\label{B21}
\end{eqnarray}
$ (\partial \tilde U)_{ij} = (\nabla_i \tilde U_{j} + \nabla_j
\tilde U_{i}) / 2 ,$ $ \, \omega_{ij} = \hat \omega_i \hat
\omega_j ,$ $ \, \hat \omega_i = \Omega_i / \Omega ,$ and
\begin{eqnarray*}
D_1(\omega) &=& \{2 \, [A_{1}^{(1)}(\omega) - A_{1}^{(1)}(0) + (q
+ 2) C_{1}^{(1)}(0)
\nonumber\\
&& + C_{1}^{(1)}(\omega)] + A_{2}^{(1)}(\omega)\} / 4 \;,
\\
D_2(\omega) &=& [2 C_{3}^{(1)}(\omega) - A_{2}^{(1)}(\omega)] / 8
\;,
\\
D_3(\omega) &=& - (1/8) A_{2}^{(1)}(\omega) \;, \, D_4(\omega) =
(1/4) C_{2}^{(1)}(\omega)  \;,
\\
D_5(\omega) &=& 2 \, [4 C_{1}^{(2)}(\omega) + C_{2}^{(2)}(\omega)
+ 7 \, C_{3}^{(2)}(\omega)] \;,
\\
D_6(\omega) &=& C_{2}^{(2)}(\omega) + 2 C_{3}^{(2)}(\omega) \;,
\end{eqnarray*}
$\omega = 8 \tau_0 \Omega .$ The functions $A_{m}^{(n)}(\omega)$
and $C_{m}^{(n)}(\omega)$ are determined in Appendix C.
Equation~(\ref{B14}) for the effective force $ \tilde{\bec{\cal
F}} = - {\nabla}_j  \tilde f_{ij}({\bf R})$ can be rewritten in
the form
\begin{eqnarray}
\tilde{\cal F}_i = \tilde{\cal F}_i^{(H)} + (\nu_{_{T}}^{(0)} \,
M_{ij} + {1 \over 6} l_0^2 \, \Omega \, S_{ij}) \Lambda_j \;,
\label{AB14}
\end{eqnarray}
where $\nu_{_{T}}^{(0)} = \tau_0 \langle {\bf u}^2 \rangle^{(0)} /
6 = l_0^2 /6 \tau_0,$ $\, {\bf \Lambda} = (\bec{\nabla} l_0^2) /
l_0^2 = (\bec{\nabla} \nu_{_{T}}^{(0)}) / \nu_{_{T}}^{(0)} ,$ and
the effective force $\tilde{\cal F}_i^{(H)}$ in a homogeneous
turbulence reads
\begin{eqnarray}
\tilde{\cal F}_i^{(H)} &=& \nu_{_{T}}^{(0)} \biggl\{ \biggl[{1
\over 2} D_1 - D_3 \biggr] \Delta \tilde U_i + D_4 \, \hat
\omega_{i} (\bec{\hat{\omega}} \cdot \bec{\nabla})^2
(\bec{\hat{\omega}} \cdot \tilde{\bf U})
\nonumber\\
&& + (D_2 + D_3) \, [(\bec{\hat{\omega}} \cdot \bec{\nabla})^2
\tilde U_i + \hat \omega_{i} \, \Delta (\bec{\hat{\omega}} \cdot
\tilde{\bf U})] \biggr\}
\nonumber\\
&& + {1 \over 6} l_0^2 \, \Omega \, \biggl[D_6 \,
(\bec{\hat{\omega}} \cdot \bec{\nabla}) [(\bec{\hat{\omega}} {\bf
\times} \bec{\nabla})_i (\bec{\hat{\omega}} \cdot \tilde{\bf U})
\nonumber\\
&&  - (\bec{\hat{\omega}} \cdot \tilde{\bf W}) \omega_{i} +
(\bec{\hat{\omega}} \cdot \bec{\nabla}) (\bec{\hat{\omega}} {\bf
\times} \tilde{\bf U})_i]
\nonumber\\
&& - {1 \over 2} D_5 \, \Delta (\bec{\hat{\omega}} {\bf \times}
\tilde{\bf U})_i \biggr] + {1 \over 12} l_0^2 \, \Omega \, D_5 \,
\nabla_i (\bec{\hat{\omega}} \cdot \tilde{\bf W})
\nonumber\\
&& + \nu_{_{T}}^{(0)} (D_2 - D_3) \nabla_i (\bec{\hat{\omega}}
\cdot \bec{\nabla}) (\bec{\hat{\omega}} \cdot \tilde{\bf U}) \;,
\label{AB15}
\end{eqnarray}
where we used the identity
\begin{eqnarray*}
\varepsilon_{ijk} \varepsilon_{lmn} &=&  \delta_{il} \delta_{jm}
\delta_{kn} + \delta_{in} \delta_{jl} \delta_{km} + \delta_{im}
\delta_{jn} \delta_{kl}
\nonumber \\
& & - \delta_{in} \delta_{jm} \delta_{kl} - \delta_{il}
\delta_{jn} \delta_{km} - \delta_{im} \delta_{jl} \delta_{kn} \; .
\end{eqnarray*}
Equation~(\ref{AB15}) for the effective force $\tilde{\cal
F}_i^{(H)}$ in a homogeneous turbulence coincides in the form with
that obtained in \cite{G03} using symmetry arguments. However, the
symmetry arguments cannot allow to determine the coefficients in
Eq.~(\ref{AB15}).

\medskip

\section{The identities used for the integration in ${\bf k}$-space}

To integrate over the angles in $ {\bf k} $--space we used the
following identities:
\begin{eqnarray}
\bar J_{ij}(a) = \int {k_{ij} \sin \theta \over 1 + a \cos^{2}
\theta} \,d\theta \,d\varphi =  \bar A_{1} \delta_{ij} + \bar
A_{2} \, \omega_{ij} \;, \label{AP1}
\end{eqnarray}
\begin{widetext}
\begin{eqnarray}
\bar J_{ijmn}(a) &=& \int {k_{ijmn} \sin \theta \over 1 + a
\cos^{2} \theta} \,d\theta \,d\varphi = \bar C_{1} (\delta_{ij}
\delta_{mn} + \delta_{im} \delta_{jn} + \delta_{in} \delta_{jm}) +
\bar C_{2} \, \omega_{ijmn}
\nonumber\\
& &+ \bar C_{3} (\delta_{ij} \omega_{mn} + \delta_{im} \omega_{jn}
+ \delta_{in} \omega_{jm} + \delta_{jm} \omega_{in} + \delta_{jn}
\omega_{im} + \delta_{mn} \omega_{ij}) \;,
\label{AP2} \\
\bar H_{ijmn}(a) &=& \int {k_{ijmn} \sin \theta \over (1 + a
\cos^{2} \theta)^{2} } \,d\theta \,d\varphi = - \biggl( {\partial
\over \partial b } \int {k_{ijmn} \sin \theta \over b + a \cos^{2}
\theta} \,d\theta \,d\varphi \biggr)_{b=1}
\nonumber\\
&=& \bar J_{ijmn}(a) + a {\partial \over \partial a} \bar
J_{ijmn}(a) \;, \label{AP4}
\end{eqnarray}
\end{widetext}
\noindent where $ \omega_{ij} = \hat \omega_{i} \hat \omega_{j} ,$
$ \quad \omega_{ijmn} = \omega_{ij} \omega_{mn} ,$ $ \, \bar A_{1}
= 5 \bar C_{1} + \bar C_{3} ,$ $ \, \bar A_{2} = \bar C_{2} + 7
\bar C_{3} ,$ and
\begin{eqnarray*}
\bar A_{1}(a) &=& {2 \pi \over a} \biggl[(a + 1) {\arctan
(\sqrt{a}) \over \sqrt{a}} - 1 \biggr] \;,
\\
\bar A_{2}(a) &=& - {2 \pi \over a} \biggl[(a + 3) {\arctan
(\sqrt{a}) \over \sqrt{a}} - 3 \biggr] \;,
\\
\bar C_{1}(a) &=& {\pi \over 2a^{2}} \biggl[(a + 1)^{2} {\arctan
(\sqrt{a}) \over \sqrt{a}} - {5 a \over 3} - 1 \biggr] \;,
\\
\bar C_{2}(a) &=& {\pi \over 2a^{2}} \biggl[(3 a^{2} + 30 a + 35)
{\arctan (\sqrt{a}) \over \sqrt{a}}
\nonumber\\
& & - {55 a \over 3} - 35 \biggr] \;,
\\
\bar C_{3}(a) &=& - {\pi \over 2a^{2}} \biggl[(a^{2} + 6 a + 5)
{\arctan (\sqrt{a}) \over \sqrt{a}} - {13 a \over 3} - 5 \biggr]
\; .
\end{eqnarray*}
In the case of $ a \ll 1 $ these functions are given by
\begin{eqnarray*}
\bar A_{1}(a) &\sim& (4 \pi / 3) [1 - (1 / 5) a]  \;, \quad \bar
A_{2}(a) \sim - (8 \pi / 15) a \;,
\\
\bar C_{1}(a) &\sim& (4 \pi / 15) [1 - (1 / 7) a]  \;, \quad \bar
C_{2}(a) \sim (32 \pi / 315) a^{2},
\\
\bar C_{3}(a) &\sim& - (8 \pi / 105) a \; .
\end{eqnarray*}
In the case of $ a \gg 1 $ these functions are given by
\begin{eqnarray*}
\bar A_{1}(a) &\sim& \pi^{2} / \sqrt{a}  \;, \quad \bar A_{2}(a)
\sim - \pi^{2} / \sqrt{a} \;,
\\
\bar C_{1}(a) &\sim& \pi^{2} / 4 \sqrt{a} - 4 \pi / 3 a \;, \quad
\bar C_{2}(a) \sim 3 \pi^{2} / 4 \sqrt{a} \;,
\\
\bar C_{3}(a) &\sim& - \pi^{2} / 4 \sqrt{a} + 8 \pi / 3 a \; .
\end{eqnarray*}
Now we calculate the following functions
\begin{eqnarray*}
A_{k}^{(p)}(\omega) &=& (6 / \pi \omega^{p+1}) \int_{0}^{\omega}
y^{p} \bar A_{k}(y^{2}) \,d y \;,
\\
C_{k}^{(p)}(\omega) &=& (6 / \pi \omega^{p+1}) \int_{0}^{\omega}
y^{p} \bar C_{k}(y^{2}) \,d y \; .
\end{eqnarray*}
The integration yields:
\begin{eqnarray*}
A_{1}^{(1)}(\omega) &=& 12 \biggl[{\arctan (\omega) \over \omega}
\biggl(1 - {1 \over \omega^{2}} \biggr)
\\
\nonumber\\
&& + {1 \over \omega^{2}}[1 - \ln(1 + \omega^{2})] \biggr] \;,
\\
A_{2}^{(1)}(\omega) &=& - 12 \biggl[{\arctan (\omega) \over
\omega} \biggl(1 - {3 \over \omega^{2}} \biggr)
\nonumber\\
&& + {1 \over \omega^{2}}[3 - 2 \ln(1 + \omega^{2})] \biggr] \;,
\\
C_{1}^{(1)}(\omega) &=& {\arctan (\omega) \over \omega} \biggl(3 -
{6 \over \omega^{2}} - {1 \over \omega^{4}} \biggr)
\nonumber\\
&& + {1 \over \omega^{2}} \biggl({17 \over 3} + {1 \over
\omega^{2}} - 4 \ln(1 + \omega^{2}) \biggr) \;,
\\
A_{1}^{(2)}(\omega) &=& 6 \biggl[{\arctan (\omega) \over \omega}
\biggl(1 + {1 \over \omega^{2}} \biggr) - {3 \over \omega^{2}} +
{2 \over \omega^{3}} S(\omega) \biggr],
\\
A_{2}^{(2)}(\omega) &=& - 6 \biggl[{\arctan (\omega) \over \omega}
\biggl(1 + {1 \over \omega^{2}} \biggr) - {7 \over \omega^{2}} +
{6 \over \omega^{3}} S(\omega) \biggr],
\\
C_{1}^{(2)}(\omega) &=& (3/2) \biggl[{\arctan (\omega) \over
\omega} \biggl(1 - {1 \over \omega^{4}} \biggr) - {13 \over 3
\omega^{2}}
\nonumber\\
&& + {1 \over \omega^{4}} + {4 \over \omega^{3}} S(\omega) \biggr]
\;,
\\
C_{2}^{(p)}(\omega) &=& A_{2}^{(p)}(\omega) - 7
A_{1}^{(p)}(\omega) + 35 C_{1}^{(p)}(\omega) \;,
\\
C_{3}^{(p)}(\omega) &=& A_{1}^{(p)}(\omega) - 5
C_{1}^{(p)}(\omega) \;,
\end{eqnarray*}
where $ S(\omega) = \int_{0}^{\omega} [\arctan (y) / y] \,d y .$
In the case of $ \omega \ll 1 $ these functions are given by
\begin{eqnarray*}
A_{1}^{(1)}(\omega) &\sim& 4 (1 - {1 \over 10} \omega^{2}) \;, \,
A_{2}^{(1)}(\omega) \sim - {4 \over 5} \omega^{2} \;,
\\
A_{1}^{(2)}(\omega) &\sim& {8 \over 3} (1 - {3 \over 25}
\omega^{2}) \;, \, A_{2}^{(2)}(\omega) \sim - {16 \over 25}
\omega^{2} \;,
\\
C_{1}^{(1)}(\omega) &\sim& {4 \over 5} (1 - {1 \over 14}
\omega^{2}) \;, \, C_{2}^{(1)}(\omega) \sim O(\omega^{4}) \;,
\\
C_{3}^{(1)}(\omega) &\sim& - {4 \over 35} \omega^{2} \;, \,
C_{1}^{(2)}(\omega) \sim {8 \over 15} (1 - {3 \over 35}
\omega^{2}) \;, \,
\\
C_{2}^{(2)}(\omega) &\sim& O(\omega^{4}) \;, \,
C_{3}^{(2)}(\omega) \sim - {16 \over 175} \omega^{2} \; .
\end{eqnarray*}

\end{document}